\documentclass[12pt]{article}

\usepackage{amsmath,amssymb}
\usepackage{graphicx}
\usepackage{caption}
\usepackage{color}
\usepackage{dcolumn}
\usepackage{bm}
\usepackage[numbers,super,comma,sort&compress]{natbib}

\definecolor{background-color}{gray}{0.98}

\title{On the status of the Born-Oppenheimer expansion in molecular systems theory}
\author{A. Yu.~Zakharov\thanks{Yaroslav-the-Wise Novgorod State University, Veliky Novgorod, 173003, Russia, e-mail: Anatoly.Zakharov@novsu.ru; A.Yu.Zakharov@gmail.com}}

\begin{document}

\maketitle

\begin{abstract}
It is shown that the adiabatic Born-Oppenheimer expansion does not satisfy the necessary condition for the applicability of perturbation theory. A simple example of an exact solution of a problem that can not be obtained from the Born-Oppenheimer expansion is given. A new version of perturbation theory for molecular systems is proposed. 
\end{abstract}

\clearpage


  \makeatletter
  \renewcommand\@biblabel[1]{#1.}
  \makeatother

\bibliographystyle{apsrev}

\renewcommand{\baselinestretch}{1.5}
\normalsize

\clearpage

\section*{\sffamily \Large INTRODUCTION} 


At present, nonrelativistic quantum theory of molecular systems is based on the Born-Oppenheimer expansion. The Hamiltonian of a system consisting of electrons and nuclei is represented in the form~\cite{BO,BH}
\begin{equation}\label{Ham}
\hat{H}\left(x, \frac{\partial}{\partial x}, X, \frac{\partial}{\partial X} \right) = \hat{H}_0\left(x, \frac{\partial}{\partial x}, X \right) + \kappa^4\,  \hat{H}_1\left( \frac{\partial}{\partial X} \right), 
\end{equation}
where $H_0\left(x, \frac{\partial}{\partial x}, X \right) $ contains the operators of kinetic energy of electrons and the total Coulomb energy of electrons and nuclei,
\begin{equation}\label{H1}
\kappa^4\, \hat{H}_1\left(\frac{\partial}{\partial X } \right) = -\sum \frac{\hbar^2}{2M}\, \frac{\partial^2}{\partial X^2} = -\kappa^4 \sum \left(\frac{M_0}{M} \right) \frac{\hbar^2}{2m}\, \frac{\partial^2}{\partial X^2} 
\end{equation}
is the operator of kinetic energy of the nuclei, small letters $(m,x)$  and capital letters $(M,X)$ relate to electrons and nuclei, respectively, $\kappa=\left( \frac{m}{M_0}\right)^{1/4} $, $M_0$ is any of the nuclear masses or their mean value. 

The Hamiltonian~(\ref{Ham}) contains a small parameter~$\kappa$ and the Born-Oppenheimer expansion is a perturbation theory in this parameter.

As is well known~\cite{Landau,Mayer}, the wave function of the total Hamiltonian~$\hat{H}$ is represented as an expansion in eigenfunctions of the unperturbed Hamiltonian~$\hat{H}_0$. This means that both perturbed and unperturbed wave functions must belong to the same Hilbert space~\cite{Kato,Yafaev}. Such a situation can be, for example, in single-particle problems. 

In the quantum many-body problem, and for the Hamiltonian~(\ref{Ham}) in particular, it is necessary that the domains of operators $\hat{H}$ and $\hat{H}_0$ be identical. The Born-Oppenheimer perturbation theory does not satisfy this requirement.

The work is organized as follows. The second section contains an example of a simple problem that contains a small parameter similar to the $\kappa$ parameter in the Schr\"{o}dinger equation for a molecular system. It is shown that this solution can not be obtained by the Born-Oppenheimer type perturbation theory. The third section contains a slightly more detailed analysis of the Born-Oppenheimer expansion and a discussion of other possible options for choosing the unperturbed Hamiltonian for molecular systems.

\section*{\sffamily \Large A COUNTEREXAMPLE TO THE BORN-OPPENHEIMER TYPE EXPANSION}

Let us consider a two-dimensional equation of elliptic type with a small parameter  $ \varepsilon > 0$ for one of the partial derivatives:
\begin{equation}\label{Lapl}
\frac{\partial^2 u}{\partial x^2} + \varepsilon\, \frac{\partial^2 u}{\partial y^2} = 0.
\end{equation}
Using the transformation of variables
\begin{equation}\label{change}
\left\lbrace 
\begin{array}{rcl}
 z & = &  x+i\frac{y}{ \sqrt{\varepsilon} } ; \\
  z^* & = &  x-i\frac{y}{ \sqrt{\varepsilon} }, 
\end{array}
\right. 
\end{equation}
the equation~(\ref{Lapl}) has the following form
\begin{equation}\label{Lapl2}
\frac{\partial^2 u}{\partial z\, \partial z^*}=0.
\end{equation}
This implies the general solution of equation~(\ref{Lapl})
\begin{equation}\label{sol}
u\left(x,y \right) = \varphi\left( x+i\frac{y}{ \sqrt{\varepsilon}} \right) + \chi\left(  x-i\frac{y}{ \sqrt{\varepsilon}}\right), 
\end{equation}
where $ \varphi(z)$ and $\chi(z^*) $ are some arbitrary twice differentiable functions.

Any particular solution of equation~(\ref{Lapl}) can be obtained from the general solution~(\ref{sol}) by using the appropriate boundary conditions.

It should be noted that both the general solution~(\ref{sol}) and the particular solutions do not allow the passage to the limit $\varepsilon\to +0 $, so the perturbation associated with the second term in~(\ref{Lapl}) is singular. In the best case, the expansions of the general and particular solutions of the equation~(\ref{Lapl}) can be the Laurent type series with respect to the parameter~$\varepsilon^{1/2}$. 
Thus, the use of perturbation theory in the small parameter $ \varepsilon$  to obtain solutions of the equation~(\ref{Lapl}) is incorrect.

The reason for this incorrectness is as follows. The small parameter~$ \varepsilon $ is at the highest derivative with respect to $y$.
If the parameter $\varepsilon\not=0 $, then the solution of the equation~(\ref{Lapl}) depends both on the equation and on the boundary conditions. As soon as the parameter $\varepsilon$ vanishes, one of the independent variables disappear in this equation. This means that there is a change in the type of the equation and the old boundary conditions become incompatible with the changed equation.

\section*{\sffamily \Large A MORE DETAILED ANALYSIS OF THE BORN-OPPENHEIMER EXPANSION}

The situation with expansion of the Schr\"{o}dinger equation solutions for the Hamiltonian~(\ref{Ham}) is completely analogous to the counterexample considered. Operators $\hat{H}$ and $\hat{H}_0$ act in different Hilbert spaces. The domain of the operator $\hat{H}$ is $L^2\left(R^{3\left(N_1 + N_2\right) } \right) $, where $N_1$ and $N_2$ are total numbers of electrons and nuclei, respectively. The domain of the operator $\hat{H}_0$ is $L^2\left(R^{3N_1} \right) $. Hilbert space  $L^2\left(R^{3\left(N_1 + N_2\right) } \right) $ contains the space $L^2\left(R^{3N_1} \right) $ as a proper subspace. Therefore, a basis in the space $L^2\left(R^{3N_1} \right) $ is not a basis in the ambient space $L^2\left(R^{3\left(N_1 + N_2\right) } \right) $. 

Thus, the use of perturbation theory with respect to the $\kappa$ parameter does not allow to get out the subspace~$L^2\left(R^{3N_1} \right) $, which is only a part of the total Hilbert space~$L^2\left(R^{3\left(N_1 + N_2\right) } \right) $.  It is well known, for perturbation theory to be applicable, it is necessary (but not sufficient) that the domains of both perturbed and unperturbed operators be identical.

The analogous situation holds in the theory of superconductivity in the framework of the BCS model. The exact solution of this model can not also be obtained by perturbation theory.

In this connection, the question arises of a suitable definition of the unperturbed Hamiltonian~$\hat{H}_0$ for molecular systems. This definition must satisfy the following requirements.
\begin{enumerate}
	\item The domain of this operator must be ~$L^2\left(R^{3\left(N_1 + N_2\right) } \right) $.
	\item The spectrum of this operator should ensure the possibility of bound states of the system already in the zeroth approximation.
\end{enumerate}
It is clear that the Hamiltonian of free particles does not satisfy these requirements. Moreover, the Hamiltonians of such type
\begin{equation}\label{bad-Ham}
\begin{array}{l}
{\displaystyle \hat{H}_0^{\left( e\right) } = - \sum_{s=1}^{N_1}\frac{\hbar^2}{2m}\Delta_s + \frac12\sum _{\substack{s,s'=1\\ s\not = s'}  }^{N_1} v^{\left( ee\right) }_{\mathrm{Coulomb}}\left( \mathbf{r}_s - \mathbf{r}_{s'}\right), }\\
{\displaystyle 
\hat{H}_0^{\left( n\right) } = - \sum_{s=1}^{N_2}\frac{\hbar^2}{2M_s}\Delta_s + \frac12\sum _{\substack{s,s'=1\\ s\not = s'}  }^{N_2} v^{\left( nn\right) }_{\mathrm{Coulomb}}\left( \mathbf{R}_s - \mathbf{R}_{s'}\right)}
\end{array}
\end{equation}
for electrons and nuclei, respectively, also do not suitable, since their spectra do not contain bound states and the corresponding systems of particles are scattered in the space.

To avoid such a situation, we add and subtract into the Hamiltonian of a molecular system the ``auxiliary potentials'' of the following type
\begin{equation}\label{add-terms}
\frac12\sum _{\substack{s,s'=1\\ s\not = s'}  }^{N_1} w\left( \mathbf{r}_s - \mathbf{r}_{s'}\right) + \frac12\sum_{\substack{s,s'=1\\ s\not = s'}  }^{N_2} W\left( \mathbf{R}_s - \mathbf{R}_{s'}\right), 
\end{equation}
where $w\left( \mathbf{r}_s - \mathbf{r}_{s'}\right)$ and $W\left( \mathbf{r}_s - \mathbf{r}_{s'}\right) $ are some functions such as potential wells that ensure the existence of bound states of both electrons and nuclei, respectively.

As a result, we have the Hamiltonian of a system consisting of electrons and nuclei to the following form
\begin{equation}\label{new-Ham}
\hat{H} = \hat{H}_0 + \hat{H}_1,
\end{equation}
where $\hat{H}_0$ is the unperturbed Hamiltonian
\begin{equation}\label{newH0}
\begin{array}{c}
{\displaystyle \hat{H}_0 = \left[ - \sum_{s=1}^{N_1}\frac{\hbar^2}{2m}\Delta_s + \frac12 \sum _{\substack{s,s'=1\\ s\not = s'}  }^{N_1} w\left( \mathbf{r}_s - \mathbf{r}_{s'}\right)\right] }\\ {\displaystyle + \left[ - \sum_{s=1}^{N_2}\frac{\hbar^2}{2M_s}\Delta_s + \frac12 \sum_{\substack{s,s'=1\\ s\not = s'}  }^{N_2} W\left( \mathbf{R}_s - \mathbf{R}_{s'}\right)\right]  }
\end{array}
\end{equation}
and $ \hat{H}_1$ is the perturbation operator
\begin{equation}\label{newH1}
\begin{array}{c}
{\displaystyle 
\hat{H}_1 = \frac12 \sum _{\substack{s,s'=1\\ s\not = s'}  }^{N_1} \left(   v^{\left( ee\right) }_{\mathrm{Coulomb}}\left( \mathbf{r}_s - \mathbf{r}_{s'}\right) -  w\left( \mathbf{r}_s - \mathbf{r}_{s'}\right) \right) }\\
{\displaystyle + \frac 12 \sum _{\substack{s,s'=1\\ s\not = s'}  }^{N_2}\left(  v^{\left(nn \right) }_{\mathrm{Coulomb}}\left( \mathbf{R}_s - \mathbf{R}_{s'}\right) - W\left( \mathbf{R}_s - \mathbf{R}_{s'}\right)\right) }\\
{\displaystyle + \sum_{s=1}^{N_1}\sum_{s'=1} ^{N_2} v^{\left(en \right) }_{\mathrm{Coulomb}} \left(\mathbf{r}_s - \mathbf{R}_{s'}\right) }.
\end{array}
\end{equation}

Note that the functions $  w\left( \mathbf{r}_s - \mathbf{r}_{s'}\right) $ and $ W\left( \mathbf{r}_s - \mathbf{r}_{s'}\right)$ are completely arbitrary and can be chosen so that the eigenfunctions and spectra of the Hamiltonian $\hat{H}_0 $ are known. As such, for example, one can use the potentials of the oscillator type
\begin{equation}\label{oscillator}
 w\left( \mathbf{r}_s - \mathbf{r}_{s'}\right) = \frac{m\omega^2 \left( \mathbf{r}_s - \mathbf{r}_{s'}\right)^2}{2}, \quad  W\left( \mathbf{R}_s - \mathbf{R}_{s'}\right) = \frac{m\Omega^2 \left( \mathbf{R}_s - \mathbf{R}_{s'}\right)^2}{2}.
\end{equation}
In this case, the operator~(\ref{newH0}) corresponds to a system of independent oscillators with free parameters $\omega$ and $\Omega$. Here there is some analogy of this approach with the method of second quantization. The free parameters $\omega$ and $\Omega$ can be determined from the condition of the highest convergence rate of perturbation theory.

However, other variants of the choice of auxiliary potentials $  w\left( \mathbf{r}_s - \mathbf{r}_{s'}\right) $ and $ W\left( \mathbf{r}_s - \mathbf{r}_{s'}\right)$  are possible. In particular, consideration of auxiliary potentials of a general type in the analytic approach is also possible. The final results ultimately do not depend on the choice of auxiliary potential, so the choice of these potentials is mainly predetermined from reasons of convenience of computation and the speed of convergence of the corresponding computational algorithms.

\section*{\sffamily \Large DISCUSSION}

The paper contains the following results.

\begin{itemize}
\item It is established that the perturbation theory method for molecular systems in the Born-Oppenheimer form does not satisfy the necessary condition for the applicability of perturbation theory. A counterexample to the Born-Oppenheimer decomposition is given. 
\item An alternative version of perturbation theory for molecular systems is proposed.
\end{itemize}

\section*{\sffamily \Large CONCLUSION}

\subsection*{\sffamily \Large ACKNOWLEDGMENTS}
This work is fulfilled by partial financial support Russian Ministry of Education and Science within the framework of the state order (Project No.3.3572.2017).

\clearpage



\begin{thebibliography}{99}

\bibitem{BO} {Born, M., Oppenheimer, R.,} Ann. der Physik, \textbf{1927}, 84, 457--484.


\bibitem{BH}  {Born, M., Huang, K., } Dynamical Theory of Crystal Lattices. Clarendon Press, 1954.


\bibitem{Landau} { Landau, L. D., Lifshitz, E. M., } Quantum Mechanics. Non-relativistic Theory. Pergamon, 1991. 


\bibitem{Mayer}    {Mayer, I.,} Simple Theorems, Proofs, and Derivations in Quantum Chemistry. Springer, 2003.


\bibitem{Kato} {Kato, T.,} Perturbation Theory of Linear Operators. Springer, 1995.


\bibitem{Yafaev} {Yafaev, D.R.,} Mathematical scattering theory: Analytic Theory. AMS, 2010.


\end{thebibliography}


\end{document}